# Raman microscopy as a tool to estimate the H content of hydrogenated amorphous carbon layers

C. Pardanaud[a], C. Martin[a], P. Roubin[a], G. Giacometti[a],
C. Hopf[b], T. Schwarz-Selinger[b], W. Jacob[b].
[a] *Aix-Marseille Université, CNRS, PIIM UMR 7345, 13397 Marseille cedex 20, France.*
[b] *MPI für Plasmaphysik, EURATOM Association, Boltzmannstr. 2, 85748 Garching, Germany.*

**Abstract**

We revisit here how Raman microscopy can be used to estimate the H content in hydrogenated amorphous carbon layers. The H content was varied from 2 at.% to 30 at.%, using heat treatments from room temperature to 1300 K and was determined independently using ion beam analysis. We examine the correlation of various Raman parameters and the consistency of their thermal evolution with thermo-desorption results. We show that the $H_D/H_G$ parameter is linear in the full range of H content and can thus be used to estimate the H content. Conversely, we show that the $m/H_G$ parameter should be used with care, first because it is sensitive to photoluminescence quenching processes and second because it is sensitive to only H bonded to $C(sp^3)$. We also identify a weak band at 860 cm$^{-1}$ attributed to H bonded to $C(sp^2)$.

## 1. Introduction

Raman spectroscopy is routinely used to characterize C-based materials. Interpreting the 1000 - 1800 cm$^{-1}$ spectral region gives information on the carbon hybridization: if samples contain sp$^2$-carbon atoms, denoted here as $C(sp^2)$, two bands dominate this spectral range, the well known G and D bands [1]. The G band is due to the bond stretching of both aromatic and aliphatic $C(sp^2)$-$C(sp^2)$ pairs, whereas the D band (sometimes accompanied by a D' band) is due to the breathing of aromatic rings. They contain information on disorder such as the size of aromatic domains, the existence of $C(sp^3)$ linked to $C(sp^2)$ and the H content [1-3]. The Raman parameters generally used to probe the structure are: the peak wavenumber and the width of the G and D bands, denoted $\Gamma_{G,D}$ and $\sigma_{G,D}$ respectively, and the peak height ratio of these two bands, $H_D/H_G$ [2]. Depending on the kind of material, they were found to be more or less correlated [4, 5]. The D band peak height was found to be correlated with the H content in graphane [6]. For a-C:H layers, an additional spectral feature has to be taken into account: a photoluminescence background which is superimposed on the Raman spectrum. The slope of this background is calculated between 800 and 2000 cm$^{-1}$ and denoted as m. The higher m, the higher is the photoluminescence intensity. Electron-hole pairs are created by the laser used to produce the Raman effect. Relaxation in a-C:H then occurs following two competitive ways: photoluminescence and non-radiative recombination. The former process arises from the radiative recombination of the electron-hole pairs trapped in band tail states whereas the latter process can be due to the existence of paramagnetic dangling bonds inside the material. As the result of the competition between these two mechanisms, photoluminescence can be more or less quenched. Passivating these bonds, for example with hydrogen, removes this quenching [7]. Passivation can be done by annealing at low temperature (T < 600 K) which allows structural reorganization to occur [8, 9]. Nevertheless, the $m/H_G$ parameter measured with a $\lambda_L = 514$ nm laser is often used to estimate the H content by comparison with an empirical linear law determined in [4] where the authors used numerous kinds of a-C:H layers. It was found that this parameter is linked to the hydrogen content of a-C:H layers by a linear relation: H/H+C = 21.7 + 16.6 $\log_{10}$ ($m/H_G$). Recently a similar linear relation has been found although with different coefficients [10]. The lowest H contents measured with this technique are H/H+C ~ 20 at.%.

We revisit here the H content determination of a-C:H layers by means of Raman microscopy. To this aim, we analyze the evolution of Raman parameters with the structural and chemical changes under heat treatment from room temperature to 1300 K, the H content thus varying from 30 at.% to 2 at.%. Ion Beam Analyses (IBA) were used to determine the H content and Thermo-Desorption Spectrometry (TDS) was used to characterize the H release.

## 2. Experimental

220 nm thick amorphous, hydrogenated carbon thin films (a-C:H) were deposited on a Si wafer (10 cm in diameter) on the driven electrode of a capacitively coupled rf plasma (13.56 MHz) in pure methane at 2 Pa and a dc self bias of -300V. Under these conditions typical hard a-C:H layer, with H/H+C ~ 0.3 at.%, density ~ 1.9 g cm$^{-3}$ (see, e. g., Table 1 in [11]) are formed. To vary the hydrogen content, the wafer was cut in several samples about 1 cm$^2$ in size. Different samples were heat treated under ultra high vacuum conditions at selected annealing temperatures for ~ 60 minutes each. The annealing temperatures were varied from 400 to 1300 K. The as deposited layer will be called hereafter AD layer and the heat treated samples will be called hereafter HT samples. A helium bombarded Highly Oriented Pyrolytic Graphite (HOPG) was also used as reference (~ 800 eV, ~ 10$^{18}$ cm$^{-2}$, see [12] for details).

Raman spectra were obtained at room temperature with a Horiba-Jobin-Yvon HR LabRAM apparatus (laser wavelength $\lambda_L$ = 514.5 nm, 100X objective, resolution ~1 cm$^{-1}$, probe spot ~ 1 µm$^2$). The laser power was chosen as P ~ 0.2 mW m$^{-2}$ to prevent damaging of the samples. A 100 second integration time was sufficient to obtain a good signal to noise ratio in the 1000 - 1800 cm$^{-1}$ range whereas 1250 seconds were needed in the 350 - 900 cm$^{-1}$ range. The main Raman parameters analyzed were the G band wavenumber, $\sigma_G$, its full-width at half-maximum, $\Gamma_G$, the relative heights of the G and D bands, $H_D/H_G$, and the $m/H_G$ ratio. A linear background was subtracted (slope m) and heights $H_D$ and $H_G$ were measured without any fitting to prevent from ambiguousness due to a model-dependent fitting procedure. $H_D$ was therefore measured at its apparent maximum, except for the AD layer and the 600 K HT sample, for which the D band maximum was not well enough defined (see Fig. 1). For these two samples $H_D$ was taken at 1370 cm$^{-1}$. Note that $\Gamma_G$ will not be discussed in what follows as it follows the same behavior as $\sigma_G$ [13].

The H/(H+C) ratio of the layers was determined by ion beam analysis. The use of a 3 MeV $^3$He ion beam under an impact angle of 75° with respect to the surface normal allowed for the simultaneous measurement of $^1$H in the sample via elastic recoil detection and of $^{12}$C via nuclear reaction analysis using the $^{12}$C($^3$He,p)$^{14}$N reaction. Only the relative changes of the H and C areal densities were determined with respect to the as-deposited sample for which H/(H+C) = 0.3 was assumed. This value had been obtained previously on non-heat-treated samples deposited under conditions almost identical to those used in this work using an absolute quantification procedure that invoked calibration samples.

Hydrogen release during heating was quantified by thermal desorption spectroscopy. An as-deposited sample was heated from room temperature to 1350 K in a quartz tube immersed in a tube furnace at a heating rate of 15 K/min and the signals of various desorbing species, especially the signals at *m/q* = 2 and 16 that mostly originate from H$_2$ and CH$_4$, were recorded by a remote quadrupole mass spectrometer, [14].

## 3. Results and discussion

*3.1 Thermal evolution of Raman spectra*

Figure 1.a displays the Raman spectra of the AD layer and the HT samples in the spectral range 1000-1800 cm$^{-1}$. The Raman spectrum of the AD layer is composed of a G band, situated at 1522 cm$^{-1}$, and a broad shoulder at ~ 1200 – 1400 cm$^{-1}$ containing the D band. Upon heating up to 825 K, the G band narrows

and blue shifts by 78 cm$^{-1}$. For higher annealing temperatures the G band position remains constant. Such a blue shift has been previously attributed to a decrease of the number of C(sp$^3$) linked to C(sp$^2$) by correlating electron energy loss [2] or X-ray photoemission [15] spectroscopy data with Raman microscopy. The D band intensity increases up to a value close to that of the G band at 1300 K. According to the amorphisation trajectory presented in [5], these evolutions reveal the organization of the material under heating, leading to larger aromatic clusters. Figure 1b displays Raman spectra of the same samples in the 350-900 cm$^{-1}$ spectral range, comparing them with that of He bombarded HOPG. For graphene or graphite, this spectral range is usually silent because of Raman selection rules. Here, three weak and broad bands situated at 400, 700 and 860 cm$^{-1}$ (called B$_1$, B$_2$ and B$_3$) are identified for the AD layer. Their intensity is very weak: less than ~ 4 % of that of the G band. B$_1$ and B$_2$ are present both for AD and He bombarded HOPG while B$_3$ is present only for AD. This suggests that B$_3$ is most probably related to C-H bonds. Consistently, its frequency fits with those of polyaromatic hydrocarbon molecules C(sp$^2$)-H bonds. Besides, B$_1$ and B$_2$ frequencies fit well with the maxima of the phonon density of states of graphite (500, 650-875 cm$^{-1}$ [16]). B$_1$ and B$_2$ evolution with heating also differs from that of B$_3$ with respect to intensity, position and shape. When heating from room temperature to 1300 K, the intensity of B$_1$ remains roughly constant, varying from 2 % to 4 % of that of G independent of the temperature. Its profile remains broad and not well defined. The intensity of B$_2$ decreases from room temperature to 775 K and then B$_2$ stays like a plateau. Conversely, the profile of B$_3$ remains well defined upon heating. Its position clearly blueshifts by 24 cm$^{-1}$ up to 875 K and then remains constant while its intensity decreases continuously and vanishes at 1300 K.

Figures 2a and 2c display the thermal evolution of $\sigma_G$ and $\sigma_{B3}$, m/H$_G$ and H$_D$/H$_G$, respectively. They are compared to the H content as determined using IBA shown in Fig. 2c. Figure 2a displays the evolution of $\sigma_G$ and $\sigma_{B3}$ and extracted from the spectra shown in figure 1a and 1b. The plateau of $\sigma_G$ above ~ 850 K means that the number of C(sp$^3$) linked to C(sp$^2$) no longer changes. The thermal evolutions of the two bands are remarkably similar: this suggests that the C(sp$^2$)-H bond is sensitive to the presence of C(sp$^3$) linked to C(sp$^2$). Figure 2b shows that m/H$_G$ increases by ~ 40% between 300 and 600 K and then decreases monotonically from 600 to 875 K where m/H$_G$ ~ 0.05. For higher temperatures the slope m is approximately zero. On the other hand, figure 2c shows that between 300 and 600 K the H content is constant (H/H+C ~ 30 at.%). The increase of m/H$_G$ is thus probably due to an increase of photoluminescence (increasing m) due to the defect passivation occurring at the first stages of heating. Therefore, m/H$_G$ cannot be simply related to the

H content in this temperature range. At higher temperature, H/H+C decreases from 30 at.% down to ~2 at.%, without reaching a plateau. Figure 2c shows that $H_D/H_G$ and H/H+C evolve in a remarkably symmetrical way, with $H_D/H_G$ increasing continuously from ~ 0.4 to ~ 1 with T increasing from 300 to 1300 K. This suggests that the two parameters are correlated and that $H_D/H_G$ can be used to probe the hydrogen content for the full range from 2 to 30 at.%. Finally, at ~ 850 K, (i) $\sigma_G$ and $\sigma_{B3}$ reach a plateau due to the vanishing of $C(sp^3)$ linked to $C(sp^2)$, (ii) $m/H_G$ goes to zero due to the vanishing of photoluminescence, and (iii), on the contrary the H content is still changing, decreasing from ~ 15 at.% at 850 K to 2 at.% at 1300 K. This shows that Raman microscopy is sensitive to the two contributions of the H decrease, i.e. the H release from $C(sp^3)$ below 850 K and the H release from $C(sp^2)$ above 850 K, in agreement with the sequential destruction of $C(sp^3)$-H and $C(sp^2)$-H groups evidenced in [17]. This also shows that photoluminescence is due to $C(sp^3)$-H bonds. Note that the thermal evolution found here for $C(sp^3)$-H is in agreement with that found in previous photoelectron spectrum studies [15]. The evolution of $\sigma_G$ shows that the reorganization of the material already starts below 600 K, i.e. before the first stages of H release from $C(sp^3)$, together with the material reorganization which leads to passivation.

*3.2 Structure and H-content dependences of Raman parameters*

Figure 3.a displays the TDS measurements at mass 2 and mass 16 of the AD layer heated with a heating ramp of 15 K min$^{-1}$. Mass 2 and mass 16 correspond to $H_2$ and $CH_4$ which are known to be the dominant released species of this type of layer [14]. These measurements are compared to normalized absolute values of derivatives (versus the annealing temperature) of the H content measured by IBA (figure 3a) and of the Raman parameters, $\sigma_G$, $H_D/H_G$ and $m/H_G$ (figure 3b), as a function of the annealing temperature. The TDS signal of $H_2$ is composed of two contributions, centered at ~ 900 K and ~ 1100 K, whereas the TDS signal of $CH_4$ has only a contribution at ~ 900 K. These two contributions have been attributed to $C(sp^3)$-H and $C(sp^2)$-H, respectively [18]. Consistently, the temperature derivative of the H/H+C signal is also composed of two contributions, centered at ~ 825 K and ~ 975 K. The shift between these TDS and IBA results depends on the way the sample is heated, as suggested in [15]. In comparable TDS investigations it was shown that the peak maxima of the TDS peaks shift to lower temperatures if lower heating ramps are used (C. Hopf et al., to be published). $H_D/H_G$ also displays these two contributions (at ~ 775 K and ~ 975 K) while $\sigma_G$ and $m/H_G$ do not. The comparison with the TDS spectrum thus strongly

supports the analysis of figure 2 (section 3.1) and supports the supposition that $H_D/H_G$ probes all the H content, i.e. both $C(sp^3)$-H and $C(sp^2)$-H, while $m/H_G$ only probes $C(sp^3)$-H. Note that the calculus of the derivative uses 50 K steps and therefore the observed shift (~ 50 K) of the Raman parameter maxima compared to that of the IBA signal has probably no physical meaning.

To summarize the presented results, figures 4a and 4b display $m/H_G$ and $H_D/H_G$ as a function of H/H+C. In logarithmic scale, $m/H_G$ is well fitted by a linear relation: H/H+C = A + B $\log_{10}(m/H_G)$, with A = 25 and B = 9. Such a linear relation (although with different slopes) was previously obtained from a large set of samples [4, 10] leading to a significant spreading of data which are indicated in figure 4a by the grey zones. IBA probes all H in the film, i.e., both $C(sp^2)$-H and $C(sp^3)$-H; while $m/H_G$ only probes $C(sp^3)$-H and this could explain the spreading of the data and the observation of different slopes, depending on the $C(sp^3)$-H / $C(sp^2)$-H ratio of the a-C:H layers investigated in Refs. 4 and 10. Note that the AD layer point is off the straight line: this is related to the specific behavior of the $m/H_G$ parameter at low temperature, which is sensitive not only to the H content but also to the existence of non-passivated dangling bonds quenching photoluminescence. For AD layers, various amounts of non-passivated bonds thus induce various m values that may explain the spreading observed for the data presented in Ref. 4. On the other hand, $H_D/H_G$ is remarkably well fitted by a linear relation (H/H+C = 0.54 - 0.53 $H_D/H_G$) from 2 at.% to 30 at.%, i.e. as well as for H bonded to $C(sp^3)$ and for H bonded to $C(sp^2)$. The D and the G bands probe only $C(sp^2)$ carbons and this linear behavior of their relative intensities common to $C(sp^2)$ and $C(sp^3)$ properties merely reveals that the aromatic organization (aromatic cluster size, linking with defects) drives both the H content, the $C(sp^3)$ defects, and most probably more generally all the non aromatic structure. For example, although it is well known for hard layers that electron properties are driven by the $C(sp^2)$ network whereas mechanical properties are driven by the $C(sp^3)$ network ([19] and references therein), these results show how the $C(sp^3)$ network is however in close connection with the $C(sp^2)$ network.

## 4. Conclusion

Heat treatments of one type of a-C:H layer from room temperature up to 1300 K were performed and the hydrogen content evolution was studied by means of Raman microscopy, ion beam analyses and thermo-desorption spectroscopy. We have analyzed the different heating stages with respect to structure and H content, and clarified how Raman parameters are correlated to the observed changes. In particular, the contributions of H bonded either to $C(sp^2)$ or to $C(sp^3)$ are emphasized and we show that Raman microscopy gives valuable information not only on structural properties but also on chemical properties. Carbon organization starts below 600 K followed by H release from $C(sp^3)$ below 850 K and then by H release from $C(sp^2)$ at higher temperature. The relation of $\log_{10}(m/H_G)$ with the H content is found to be linear from 15 at.% to 30 at.%, extending this relation to lower values of the H content than in previous studies. We emphasize here the limits of this relation first because the $m/H_G$ signal (photoluminescence) is sensitive to defect passivation and second because $m/H_G$ is due only to H bonded to $C(sp^3)$. These two effects can lead to a misinterpretation of the H content. A linear relation is found between $H_D/H_G$ and the total H content from 2 at.% to 30 at. %, due to H bonded to either $C(sp^2)$ or $C(sp^3)$. This shows that for this type of layers $H_D/H_G$ can be used to probe the total hydrogen content. For other hard a-C:H films with comparable initial structure, this statement may also be valid. This also reveals that the whole structure of these amorphous carbon layers remains strongly driven by the aromatic network. To better understand the different processes possibly involved such as H migration, $H_2$ and $CH_4$ formation, carbon organization, their kinetics at various temperatures are currently under study.

## Acknowledgments

We acknowledge the Euratom-CEA association, the EFDA European Task Force on Plasma Wall Interactions, the Fédération de Recherche FR-FCM, the French agency ANR (ANR-06-BLAN-0008 contract) and the PACA Region (FORMICAT project) for financial support.

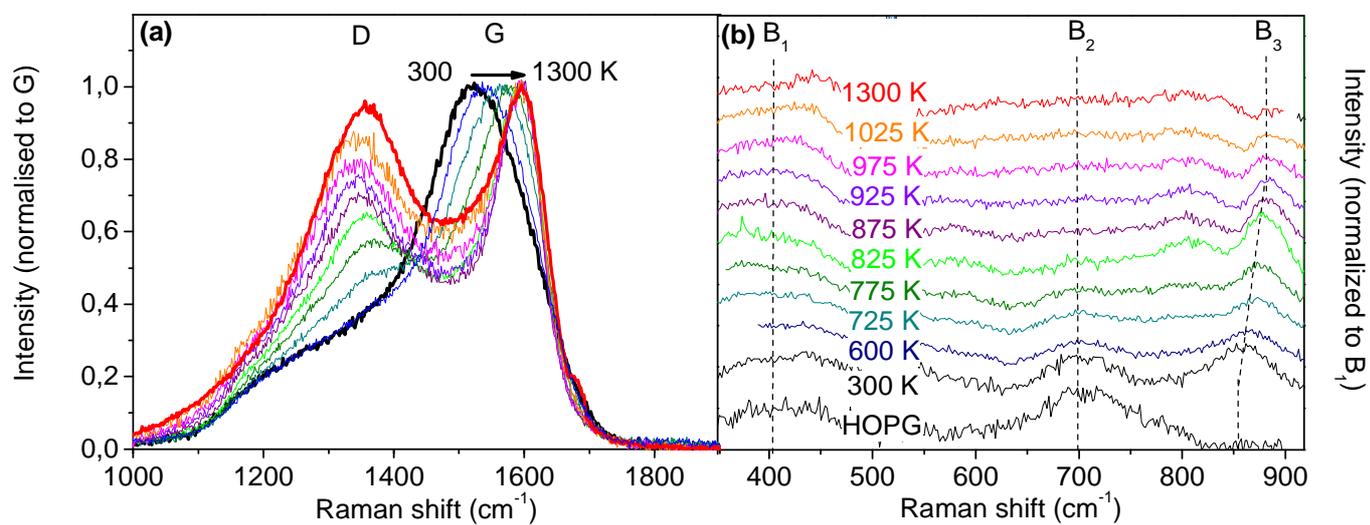

Figure 1. Raman spectra of a plasma-deposited hard amorphous carbon layer, heat treated from room temperature to 1300 K (a) spectral range 1000 - 1800 cm$^{-1}$ (the base line was subtracted and spectra normalised) (b) spectral range 350-900 cm$^{-1}$. The spectrum at the bottom corresponds to He-bombarded HOPG (see text for details).

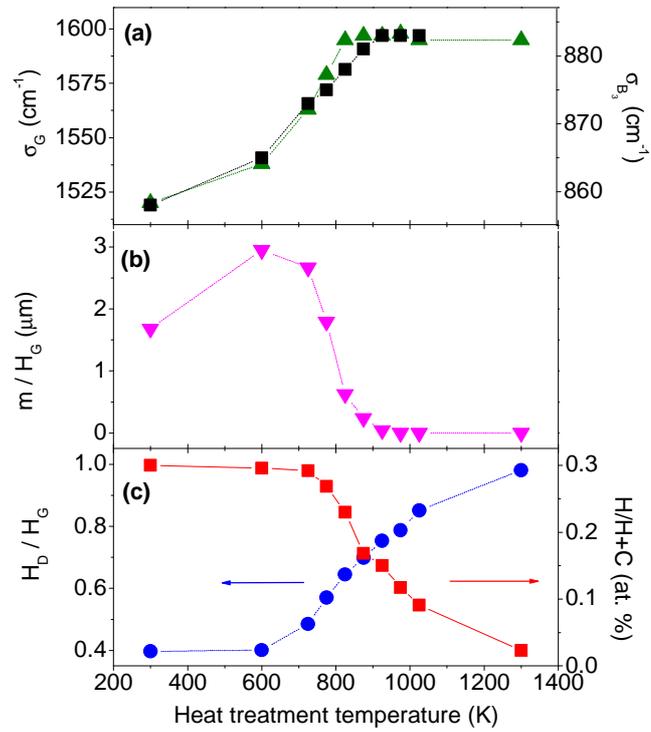

Figure 2. Evolution of Raman parameters versus the heat treatment temperature: (a) $\sigma_G$ (up triangle) and $\sigma_{B3}$ (square) (b) m/$H_G$ (down triangle) and (c) $H_D/H_G$ (circle) compared to the H content determined by IBA (square).

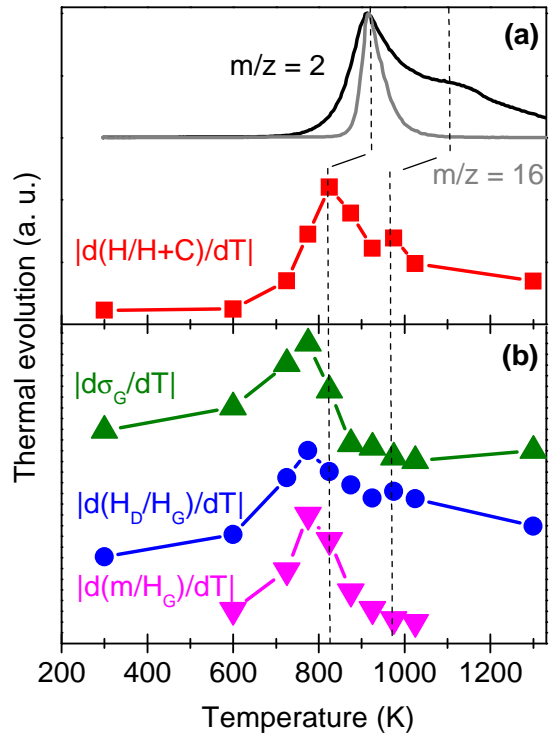

Figure 3. AD and HT layers: (a) Thermodesorption spectrum of the AD layer and normaliSed absolute value of the derivative of the IBA thermal evolution of figure 2.c (b) normalised absolute values of the derivatives of the Raman parameter thermal evolutions of figure 2.a, 2.b and 2.c.

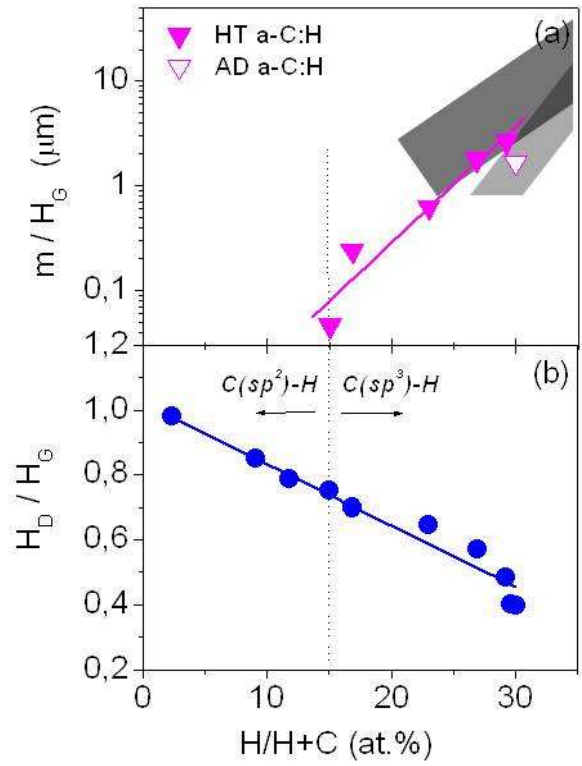

Figure 4. H content estimation using Raman parameters: (a) $m/H_G$ and (b) $H_D/H_G$ as a function of H/H+C. Dark and light grey zones correspond to previously published data (respectively [4] and [10]).